\newcommand{\be}{\begin{equation}}
\newcommand{\ee}{\end{equation}}
\newcommand{\bea}{\begin{eqnarray}}
\newcommand{\eea}{\end{eqnarray}}
\newcommand{\beann}{\begin{eqnarray*}}
\newcommand{\eeann}{\end{eqnarray*}}
\newcommand{\beasn}{\begin{sneqnarray}}
\newcommand{\eeasn}{\end{sneqnarray}}
\newcommand{\ba}{\begin{array}}
\newcommand{\ea}{\end{array}}
\newcommand{\nn}{\nonumber}
\newcommand{\Appendix}[1]%
    {\renewcommand{\thesection}{Appendix~\Alph{section}}%
     \section{#1}%
     \renewcommand{\thesection}{\Alph{section}} }

\catcode`@=11
\def\secteqno{\@addtoreset{equation}{section}%
\def\theequation{\thesection.\arabic{equation}}}
\def\endsecteqno{\def\theequation{\@ifundefined{chapter}%
{\arabic{equation}}{\thechapter.\arabic{equation}}}}
\newcounter{subequation}
\def\thesubequation{\alph{subequation}}
\def\sneqnarray{\stepcounter{equation}\let\@currentlabel=\theequation
\setcounter{subequation}{1}
\def\@eqnnum{{\rm (\theequation\thesubequation)}}
\global\@eqcnt\z@\tabskip\@centering\let\\=\@eqncr\let\@@eqncr=\@@sneqncr
$$\halign to \displaywidth\bgroup\@eqnsel\hskip\@centering
 $\displaystyle\tabskip\z@{##}$&\global\@eqcnt\@ne
 \hskip 2\arraycolsep \hfil${##}$\hfil
 &\global\@eqcnt\tw@ \hskip 2\arraycolsep
$\displaystyle\tabskip\z@{##}$\hfil
  \tabskip\@centering&\llap{##}\tabskip\z@\cr}
\def\endsneqnarray{\@@sneqncr\egroup $$\global\@ignoretrue}
\def\@@sneqncr{\let\@tempa\relax
   \ifcase\@eqcnt \def\@tempa{& & &}\or \def\@tempa{& &}
   \else \def\@tempa{&}\fi
     \@tempa \if@eqnsw\@eqnnum\stepcounter{subequation}\fi
     \global\@eqnswtrue\global\@eqcnt\z@\cr}
\def\nobiblabels{\def\@lbibitem[##1]##2{\@bibitem{##2}}}
\catcode`@=12

\def\a{\alpha}  \def\b{\beta} \def\g{\gamma} 
\def\d{\delta} \def\D{\Delta} \def\e{\epsilon}
 \def\th{\theta}  
 \def\l{\lambda}  \def\m{\mu} 
  \def\p{\pi}  
\def\s{\sigma}   
  \def\x{\chi}  
\def\o{\omega}  

\newcommand{\PRL}[3]{{\sl Phys. Rev. Lett.} {\bf #1} (19#2) {#3}}
\newcommand{\ZFP}[3]{{\sl Z. Physik} {\bf #1} (19#2) {#3}}
\newcommand{\PR}[3]{{\sl Phys. Rev.} {\bf #1} (19#2) {#3}}

\newcommand{\IJMP}[3]{{\sl Int. J. Mod. Phys.} {\bf #1} (19#2) {#3}}

\def\zb{\bar{z}}

\def\apo{a^+_0}      \def\ap3{a^+_3}
\def\amo{a^-_0}      \def\am3{a^-_3}
         \def\b3{a^3_3}

\def\apm{a^+_{\mu}}
\def\amm{a^-_{\mu}}
\def\bm{a^3_{\mu}}

\def\dpzz{d^+_{zz}}   \def\dpzbzb{d^+_{\zb\zb}}
\def\dmzz{d^-_{zz}}   \def\dmzbzb{d^-_{\zb\zb}}
\def\ezz{d^3_{zz}}    \def\ezbzb{d^3_{\zb\zb}}

\def\dpz3{d^+_{3z}}   \def\dpzb3{d^+_{3\zb}}
\def\dmz3{d^-_{3z}}   \def\dmzb3{d^-_{3\zb}}
\def\ez3{d^3_{3z}}    \def\ezb3{d^3_{3\zb}}

\def\dpmn{d^+_{pq}}
\def\dmmn{d^-_{pq}}
\def\emn{d^3_{pq}}

\def\mpp{m^{++}}
\def\mmm{m^{--}}
\def\mm3{m^{-3}}
\def\mp3{m^{+3}}
\def\mmp{m^{-+}}
\def\md3{m^{33}}

\def\hd{h^{\dagger}}

\def\Sv{{\bf S}}
\def\Dv{{\bf D}}

\def\xv{{\bf x}}

\def\Dv{{\bf D}}
\def\Ev{{\bf E}}
\def\Bv{{\bf B}}
\def\Hv{{\bf H}}
\def\Mcalv{{\bf {\cal M}}}
\def\Pcalv{{\bf {\cal P}}}

\def\pa{\partial} \def\da{\dagger} 

\documentstyle[12pt]{article}

\secteqno
\begin{document}

%%%%%%%%%%%%%%%%%%%%%%%%TITLE%%%%%%%%%%%%%%%%%%%%%%%%%%%%%%%%%%%%%%%%%%%%

\title{{\bf Spin Wave Mediated Non-reciprocal Effects \\ in \\
            Antiferromagnets }}
\author{{\Large {\sl Jos\'e Mar\'{\i}a Rom\'an}
               \ and \  {\sl Joan Soto}}\\
        \small{\it{Departament d'Estructura i Constituents
               de la Mat\`eria}}\\
        \small{\it{and}}\\
        \small{\it{Institut de F\'\i sica d'Altes Energies}}\\
        \small{\it{Universitat de Barcelona}}\\
        \small{\it{Diagonal, 647}}\\
        \small{\it{E-08028 Barcelona, Catalonia, Spain.}}\\
        {\it e-mails:} \small{roman@ecm.ub.es, soto@ecm.ub.es} }
\date{\today}

\maketitle

\thispagestyle{empty}

\begin{abstract}
By using an effective field theory for the electromagnetic interaction
of spin waves, we show that, in certain antiferromagnets, the latter
induce non-reciprocal effects in the microwave region, which should be
observable in the
second harmonic generation and produce gyrotropic birefringency.
 We calculate
the various
(non-linear) susceptibilities in terms of a few parameters the order
of magnitude of which is under control.

\end{abstract}
\bigskip
PACS: 75.30.Ds, 75.30.Gw, 75.50.Ee, 78.70.Gq, 42.65.k 
%12.39.Hg, .20.Gd

\vfill
\vbox{
\hfill{cond-mat/9709299}\null\par
\hfill{UB-ECM-PF 97/24}\null\par}

\newpage

%%%%%%%%%%%%%%%%%%%%%%%%%%%%%%%%%%%%%%%%%%%%%%%%%%%%%%%%%%%%%%%%%%%%%%%%%
%%%%%%%%%%%%%%%%%%%%%% INTRODUCTION %%%%%%%%%%%%%%%%%%%%%%%%%%%%%%%%%%%%%%

\section{Introduction}
\indent

The response of magnetic materials to electromagnetic fields gives
rise to a reach
variety of interesting phenomena \cite{Burns}. In particular,
non-reciprocal optical effects in
antiferromagnets have received considerable attention during the last
years \cite{Valenti,experimental}.
The possible existence of certain phenomena, like second harmonic
generation (SHG) or
gyrotropic birefringency (GB), is dictated by the magnetic group of the
given material,
which may (or may not) allow suitable (non-linear) susceptibilities to
be different from zero \cite{Birss,theoretical}.
The optical wavelengths induce atomic transitions which provide a
potential
microscopic mechanism to obtain non-vanishing susceptibilities.
Indeed, this is the
case for the observed non-reciprocal effects in $Cr_2O_3$
\cite{Valenti,experimental}. However, alternative
mechanisms to produce such effects cannot be ruled out {\it a priori}, 
and may even become dominant at certain wavelengths. It is our aim to 
demonstrate that this is
indeed the case for certain antiferromagnets when the electromagnetic
fields are in the microwave region. This region is very sensible to
collective magnetic effects which makes a field theoretical description
appropriate.

The low temperature low energy properties of antiferromagnets (with
spontaneous
staggered magnetization) are dominated by spin waves. The spin wave
dynamics at low
momenta and energy is very much constrained by group theoretical
considerations \cite{Has-Nie}.
The symmetry breaking pattern $SU(2)\rightarrow U(1)$
tells us that the spin waves
must transform under a non-linear realisation of $SU(2)$ \cite{Coleman}.
In addition to that the space group and time reversal 
must be respected by the dynamics.
The continuum approach 
ensures that it is enough to consider the rotational part of the space group, 
namely the point group, and the primitive translations as well as time
reversal. A systematic description of the spin wave dynamics
fully exploiting the above group
theoretical constraints has been provided in \cite{Roman}.

In this paper we apply the general framework described in \cite{Roman}
to work out the electromagnetic response of certain antiferromagnets
in the microwave region. We have
chosen a crystal with no primitive translations mapping points with opposite
magnetisations. The point group is taken to be
$\bar 3 m$, but repeating the analysis for any other point group
is straightforward. This choice is motivated by the $Cr_2O_3$ crystal,
which shows interesting non-reciprocal effects in the optical region
as mention before. We have calculated the linear and non-linear electric
and magnetic susceptibilities, which turn out to depend non-trivially on 
the frequency of the incoming radiation.
In particular non-reciprocal phenomena in the SHG as well as the GB are 
predicted to occur. These results apply to any antiferromagnet (with 
spontaneous staggered magnetisation) of arbitrary spin and crystal point 
group $\bar3m$ such that the magnetic ions lie on the $z$-axis and  
no primitive translation mapping points with
opposite magnetisation exists.

\newpage

In order to simplify the notation we will take $\hbar = c = 1$, which
lead to a relativistic notation. So $x =(t,\xv)$ and $q =(\o,{\bf k})$.
Subindices $\m = 0,1,2,3$, where the first
one represents the time component. Furthermore we work with holomorphic
coordenates $z = x + iy$ and $\zb = x - iy$.
We distribute the paper as follows. In section 2 we briefly review some basic
aspects of electromagnetic wave propagation in media in order to explain the 
appearance of non-reciprocal effects in SHG as well as GB. In section~3 we quickly 
review the framework described in full detail in \cite{Roman}. In section~4 we
present the effective lagrangian. In section~5 we work out the effective
action which describes the response to the electromagnetic field and
give the (non-linear) electric and magnetic susceptibilities. Section~6
is devoted to a discussion. Finally, in the appendix we list all the terms
which are not displayed sections~4~and~5 in order to make the presentation simpler.

%\end{document}

%%%%%%%%%%%%%%%%%% WAVE PROPAGATION %%%%%%%%%%%%%%%%%%%%%%%%%%%%%%%%%

\section{Electromagnetic Waves Propagation in Media} 
\indent

It is the aim of this section to briefly review some features of electromagnetic 
wave propagation in media, which are relevant for the rest of the paper. 
In particular the phenomena of gyrotropic birefringence (GB) and second harmonic 
generation (SHG), in connection with non-reciprocal effects, which arise due to 
time reversal violation in the medium.

\medskip

Let us recall the Maxwell equations in insulating and chargeless media
\bea
& & {\bf \nabla D} = 0  \nn \\
& & {\bf \nabla} \times \Hv = \pa_0 \Dv  \nn  \\
& & {\bf \nabla} \times \Ev = -\pa_0 \Bv  \\
& & {\bf \nabla B} = 0,  \nn 
\eea
which are to be supplemented with the constitutive equations
\be
\ba{rcl}
\Dv & = & \Ev + \Pcalv \nn \\
\Hv & = & \Bv - \Mcalv,
\ea
\label{constitutive}
\ee
where $\Pcalv$ and $\Mcalv$ are the electric and magnetic response of the medium 
respectively. $\Pcalv$ and $\Mcalv$ are functionals of the electric and magnetic 
fields and, of course, depend on the physical properties of the medium. From the 
two equations above we obtain 
\be
{\bf \nabla} \times {\bf \nabla} \times \Ev + \pa_0^2 \Ev = 
        - \left(\pa_0^2 \Pcalv + {\bf \nabla} \times \pa_0 \Mcalv \right),
\label{electric}
\ee
which is going to be the basic equation in our discussion. Once $\Pcalv$ and 
$\Mcalv$ are given this equation describes the propagation of electromagnetic 
waves in the medium.

\medskip

For definiteness, consider first the linear response of a homogeneous medium to
electric fields only. In this case, the most general
form for the electric response is \cite{Agranovich}
\be
{\cal P}^a(z) = \int{dx \chi^{ab} (z-x) E^b(x)},
\label{linear}
\ee
where $x$ and $z$ are space-time vectors. The tensor $\chi^{ab} (z-x)$ depends not
only on time, but also on space coordinates, which caracterises the spatial 
dispersive medium. We shall make the standard assumption that it varies slowly 
over the medium. This is equivalent to an 
expansion of the tensor $\chi^{ab} (\o,{\bf k})$ in powers of ${\bf k}$ 
in momentum space.  We obtain
\be
{\cal P}^a(t,{\bf z}) = \int{dt' \chi^{ab}(t-t') E^b(t',{\bf z})}
                + \int{dt' \g^{abc}(t-t')\pa_c E^b(t',{\bf z})} + \cdots.
\label{linear_spatial}
\ee
The first and second term in the electric response are the polarisation and the
quadrupolar moment respectively. Notice that the tensors associated to each order of 
the multipole expansion depend only on the frequency of the electromagnetic wave.

Let us see next how some terms in the multipole expansion of both the electric
and magnetic responses give rise to qualitatively new observable effects.
Consider (\ref{linear_spatial}) together with the leading term in the multipole 
expansion of the magnetic response (magnetoelectric term)
\be
\ba{rcl}
\Pcalv & = & \chi^{ab} E^b + \g^{abc}\pa_c E^b  \\
\Mcalv & = & \alpha^{ab} E^b
\ea
\ee
Upon substituting these expressions in (\ref{electric}), the quadrupolar 
and magnetoelectric term give rise to the 
so-called gyrotropic birefringence as we shall show next. Consider the plane wave
solution, \mbox{$E^a(x) = E^a(q)e^{-iqx} + E^a(-q)e^{iqx}$},
of the equation (\ref{electric}). Then $E^b (q)$ fulfills
\be
\left[ n^2 \d^{ab} - \left[ \e^{ab} - n^c \left(\e^{acd} \alpha^{db} -
                i\o \g^{abc} \right) \right] - n^a n^b \right] E^b(q) = 0,
\label{momentum}
\ee
where $n^a \equiv k^a/\o$ gives the propagation direction and its modulus 
the refraction index (recall that $q =(\o,{\bf k})$). Suppose first that the 
quadrupolar moment, $\g^{abc}$, and the magnetoelectric term, $\a^{bd}$, are zero.
Non trivial solutions to this equation arise from the condition that the 
determinant of the matrix on which the electric field acts vanishes.
The anisotropy of the permitivity tensor $\e^{ab}$
is generally responsible for this condition to yield two values of the refraction 
index for each propagation direction, which is known as birefringence. Namely,
two different plane waves propagate in each direction with two different
polarisations and two different velocities \cite{Born}. If the quadrupolar and
magnetoelectric terms are restored, they enter eq.~(\ref{momentum})
through an effective permitivity tensor, 
\mbox{$\e^{ab} - n^c \left(\e^{acd} \alpha^{db} - i\o \g^{abc} \right) $}, 
with a li\-ne\-ar dependence on the propagation direction, which is known as 
gyrotropic birefringence \cite{theoretical}. In particular, the equations
governing the propagation in directions ${\bf n}$ and $-{\bf n}$ are different, 
which implies that the GB is a non-reciprocal effect, since these propagation 
directions are related by time reversal. 

\medskip

Let us consider next the non-linear response of the system to electric fields. 
Thus we have to add to (\ref{linear}) new quadratic, cubic, \ldots, terms 
\cite{Shen}
\be
{\cal P}^a(z) = \int{dx \chi^{ab} (z-x) E^b(x)} + 
                \int{dx dy \chi^{abc}(z-x,z-y) E^b(x) E^c(y)} + \cdots.
\ee

In particular, the quadratic term in the above equation leads to the appearance of
second harmonic generation. Whenever exists in the field $E^a(x)$ a contribution 
of frequency $\o$ the electric response will have in addition two contributions, 
one of zero frequency and another one of frequency $2 \o$. Then, the electric 
response will be, in general, a superposition of plane waves with frequencies 
multiple of $\o$
\be
{\cal P}^a(z) = {\cal P}_0^a(z) + {\cal P}_q^a(z) + {\cal P}_{2q}^a(z) + \cdots,
\ee
leading, in turn, to a similar superposition for the electric field 
%when we consider
solution of
eq.~(\ref{electric}). The contribution to SHG comes from ${\cal P}_{2q}^a(z)$ in 
the expression above, which to lowest order can be written as
\be
\ba{c}
{\cal P}_{2q}^a(z) = P^a(2q) e^{-iqz} +  P^a(-2q) e^{iqz}  \\
                                                           \\      
           P^a(2q) = \chi^{abc}(q,q) E^b(q) E^c(q).
\ea
\ee

In order to be more explicit consider the non-linear expressions for the electric 
and magnetic responses,
\be
\ba{rcl}
{\cal P}^a & = & \chi^{ab} E^b  + \chi^{abc} E^b E^c  \\
{\cal M}^a & = & \m^{abc} E^b E^c,
\ea  
\ee
Once they are introduced in (\ref{electric}) we have a complicated non-linear 
equation. However, the non-linear terms are usually small. Therefore, if we pass 
the linear part of the response to the l.h.s.~we can calculate the electric field 
solution perturbatively, ${\bf E} = {\bf E}_{(0)} + {\bf E}_{(1)} + \cdots $. 
${\bf E}_{(0)}$ is the solution of the homogeneous equation (eq.~(\ref{momentum}) 
without the quadrupolar and magnetoelectric terms), i.e., a monochromatic plane 
wave of frequency $\o$. Then ${\bf E}_{(1)}$ follows from the equation
\be
{\bar \o}^2 \left[ {\bar n}^2 \delta^{ab} - \e^{ab} + {\bar n}^a{\bar n}^b \right]
E_{(1)}^b ({\bar q}) e^{-i{\bar q} x} =
- (2\o)^2 \left[\chi^{abc} + n^d \e^{ade} \m^{ebc} \right] E_{(0)}^b (q) 
       E_{(0)}^c (q) e^{-i2qx}.
\ee

It is clear that the solution of this equation requires in the l.h.s.~an electric 
field of frequency ${\bar \o} = 2\o$. This is called second harmonic generation. 
Notice, moreover, that the second term in the r.h.s.~depends linearly on the 
direction of the wave number. Then when this term is non-vanishing we have 
non-reciprocal effects in the second harmonic generation.

\medskip

In the discussion above we have presented the simplest situations which lead to
non-reciprocal effects. In magnetic materials, as the one we are interested in,
$\Pcalv$ and $\Mcalv$ depend both on the electric and magnetic fields. 
In this case, since ${\bf B} = {\bf n} \times {\bf E}$, the non-reciprocal effects
can be obtained from terms depending on the magnetic field in both the electric and
magnetic responses \cite{Pershan,Graham}.
Furthermore, the (generalised) susceptibilities are constrained by
the magnetic points group of the crystal. Since we are considering GB and SHG, 
which are dynamical effects, only the elements without the time reversal operator 
of the magnetic group are to be considered \cite{Valenti,Birss}. For $Cr_2O_3$ 
with the spins aligned in the third direction this is the 32 group. The allowed 
linear susceptibilities (relevant for the GB) are
\bea
P^z & = & \x_E^{z\zb}E^z + \x_B^{z\zb}B^z   \nonumber  \\
P^3 & = & \x_E^{33}E^3 + \x_B^{33}B^3   \nonumber  \\
    &   &                                             \label{pm_GB}   \\
M^z & = & \g_E^{z\zb}E^z + \g_B^{z\zb}B^z  \nonumber \\
M^3 & = & \g_E^{33}E^3 + \g_B^{33}B^3,   \nonumber
\eea
and the bilinear ones (relevant for the SHG)
\bea
P^z & = & \x_{EE}^{zzz}E^{\zb}E^{\zb} + 2\x_{EE}^{z\zb 3}E^zE^3 \nn \\
& & \mbox{} + \x_{EB}^{zzz}E^{\zb}B^{\zb}
            + \x_{EB}^{z\zb 3}E^zB^3 + \x_{EB}^{z3\zb}E^3B^z \nn \\
& & \mbox{} + \x_{BB}^{zzz}B^{\zb}B^{\zb} + 2\x_{BB}^{z\zb 3}B^zB^3
                                                                \nn \\
P^3 & = & 2\x_{EE}^{3\zb z}E^zE^{\zb} +
          \x_{EB}^{3\zb z}(E^zB^{\zb} - E^{\zb}B^z) +
          2\x_{BB}^{3\zb z}B^zB^{\zb}                   \nn \\
    &   &                                                      \label{pm_SHG} \\
M^z & = & \g_{EE}^{zzz}E^{\zb}E^{\zb} + 2\g_{EE}^{z\zb 3}E^zE^3 \nn \\
& & \mbox{} + \g_{EB}^{zzz}E^{\zb}B^{\zb}
            + \g_{EB}^{z\zb 3}E^zB^3 + \g_{EB}^{z3\zb}E^3B^z \nn \\
& & \mbox{} + \g_{BB}^{zzz}B^{\zb}B^{\zb} + 2\g_{BB}^{z\zb 3}B^zB^3
                                                                \nn \\
M^3 & = & 2\g_{EE}^{3\zb z}E^zE^{\zb} +
          \g_{EB}^{3\zb z}(E^zB^{\zb} - E^{\zb}B^z) +
          2\g_{BB}^{3\zb z}B^zB^{\zb}.                   \nn
\eea

In the remaining sections we shall calculate the contributions to the generalised
susceptibilities above due to the spin wave dynamics. We will start with a local
effective lagrangian describing the interaction between spin waves and 
electromagnetic fields. Upon integrating out the spin waves we obtain a non-local 
effective action for the electromagnetic fields, which is equivalent to having a 
free energy \cite{Pershan}, taking into account that $L_{int} = - H_{int}$. 
The electric and magnetic response, and hence all the (generalised) 
susceptibilities, can be easily obtained as follows

\be
P^a = { \d S_{{\it eff}} \over \d E^a }
\qquad ,\qquad
M^a = { \d S_{{\it eff}} \over \d B^a }.
\label{pm_der}
\ee

%%%%%%%%%%%%%%%%%% BUILDING BLOCKS %%%%%%%%%%%%%%%%%%%%%%%%%%%%%%%%%%%

\section{Building Blocks}
\indent

In this section we present the basic building blocks in the construction of
an effective lagrangian for the interaction between the spin waves and
electromagnetic fields, and their transformations under the relevant symmetries.
The method we follow was thoroughly des\-cribed in a previous article \cite{Roman}.
Here we shall only give a brief overview of it.

As it was mentioned in the introduction the spin waves are the lowest lying
excited states of the antiferromagnetic ground state associated to the
spontaneous symmetry breaking $SU(2) \rightarrow U(1)$. This tells us
that the associated field, $U(x)$, is an element of the
coset space $SU(2) / U(1)$ \cite{Coleman}, which transforms under
$SU(2)$ as follows:
\be
U(x) \rightarrow gU(x)\hd(g,U),
\label{su2transf}
\ee
where $g \in SU(2)$ and $h \in U(1)$ is a local ($U(x)$ dependent)
element
which restores $gU(x)$ to the coset space. If the
alignment direction of the local spin is the third direction $U(x)$ can
be written as
\be
U(x) = \exp{\left\{ {i\sqrt{2} \over f_{\pi} }
               \Big[\pi^1(x)S^1 + \pi^2(x)S^2\Big] \right\}},
\label{spinwaves}
\ee
where $\pi^{i}(x)$ are the spin wave fields. These fields in the complex
representation have the form $\pi^{\pm} = (\pi^1 \pm i\pi^2)/\sqrt{2}$
and the generators are written as $S_{\pm} = S^1 \pm iS^2$.

In addition to the continuous $SU(2)$ transformations the action must be
invariant under the space-time transformations. In our case we take the
$Cr_2 O_3$ as the underlying crystal in which the spin waves propagate.
$Cr_2 O_3$ enjoys the crystallographic point group $\bar3m$. The
transformation properties of the $U(x)$ field under the
$\bar3m \otimes T$ elements are
\be
   \ba{rccl}
   C^+_{3z}: & U(x)  &  \rightarrow  &  g_3U(x)\hd_3   \\
   I:        & U(x)  &  \rightarrow  &  U(x)C\hd_I    \\
   \s_y:     & U(x)  &  \rightarrow  &  g_2U(x)\hd_2   \\
   T:        & U(x)  &  \rightarrow  &  U(x)C\hd_t
   \ea
\qquad , \qquad
   \ba{l}
   C = e^{-i \pi S^2} \\
   C^{\da} = - C.
   \ea
\label{spacetransf}
\ee
The nontrivial transformation under $I$ is due to the fact that this
particular
transformation maps points with opposite local magnetisation in the
antiferromagnetic ground state. The primitive translations act trivially on $U(x)$ 
and have not been displayed. 

%\newpage

The spin-orbit is an important interaction which produces a gap in the
spectrum of the spin waves because it breaks
explicitly the $SU(2)$ symmetry. The breaking part is given by some
additional terms in the Heisenberg hamiltonian \cite{Moriya},
\be
H = \sum_{<i,j>} J_{ij}\Sv_i \Sv_j +
    \sum_{<i,j>} \Dv_{ij}(\Sv_i \times \Sv_j) +
    \sum_{<i,j>} M_{ij}^{ab} S^{a}_i S^b_j,
\label{moriya}
\ee
where the tensors $D_{ij}^a$ and $M_{ij}^{ab}$ break the
$SU(2)$ symmetry. The order of magnitude of such tensors is
$D^a \sim (\D g /g)J$ and $M^{ab} \sim (\D g /g)^2J$, where for
$Cr_2 O_3$, $\D g \sim 10^{-2} g$ \cite{LB}.
In order to introduce them in the effective theory we take their
local limit and promote them to sources with proper transformations
under $SU(2)$. By combining these sources with the $SU(2)$ generators we
obtain objects which transform covariantly under SU(2),
\be
   \ba{rcl}
   D_{pq} \equiv D_{pq}^aS^a  & \rightarrow  & gD_{pq}g^{\da}  \\
   M \equiv M^{ab}(S^a \otimes S^b + S^b \otimes S^a)   & \rightarrow  &
          (g \otimes g)M(g^{\da} \otimes g^{\da}).
   \ea
\ee

Finally they must be fixed to their more general form compatible with
the point group symmetry, namely,
\bea
 & \ba{l}
      \ba{l}
      D_{zz} = D_{zz}^- S_+         \\
      D_{\zb\zb} = D_{\zb\zb}^+ S_-
      \ea
   \ \ \ \ \ \ \ \ \ D_{zz}^- = -D_{\zb\zb}^+  \\
                                            \\
      \ba{l}
      D_{3z} = D_{3z}^+ S_-        \\
      D_{3\zb} = D_{3\zb}^- S_+
      \ea
   \ \ \ \ \ \ \ \ \ D_{3z}^+ = -D_{3\zb}^-
   \ea &                         \nonumber  \\
 &   &   \\
 & M = M^{-+}(S_+ \otimes S_- + S_- \otimes S_+) +
       M^{33}(S^3 \otimes S^3). & \nonumber
\eea

Therefore the objects from which we construct our theory are the spin
waves given by $U(x)$, the derivatives, $\pa_{\m}$, and the spin-orbit
tensors, $D_{ij}^a$ and $M_{ij}^{ab}$. Let us arrange them in a simple
form which provides us elements with easier transformations properties
under $SU(2)$
\bea
U^{\dagger}(x)i\partial_{\mu}U(x) & = & \amm(x)S_+ + \apm(x)S_- +
                                        \bm(x)S^3      \nonumber  \\
U^{\da}(x) D_{pq} U(x) & = & \dmmn(x)S_+ + \dpmn(x)S_- +
                               \emn(x)S^3              \nonumber  \\
\left(U^{\da}(x) \otimes U^{\da}(x)\right) M
          \Bigl( U(x) \otimes U(x) \Bigr) & = &
                           \mmm(x)(S_+ \otimes S_+)   \nonumber  \\
       & + & \mpp(x)(S_- \otimes S_-)  \nonumber \\
       & + & \md3(x)(S^3 \otimes S^3)            \\
       & + & \mmp(x)(S_+ \otimes S_- + S_- \otimes S_+) \nonumber \\
       & + & \mm3(x)(S_+ \otimes S^3 + S^3 \otimes S_+) \nonumber \\
       & + & \mp3(x)(S_- \otimes S^3 + S^3 \otimes S_-). \nonumber
\eea

%\newpage

 From (\ref{su2transf}) the transformation properties under $SU(2)$ for
the coefficients of the generators are
\beasn
 & \ba{ccl}
   \amm(x) & \rightarrow & e^{i\th(x)}\amm(x)            \\
   \bm(x)  & \rightarrow & \bm(x) + \partial_{\m}\th(x)
   \ea  &                        \label{asu2transf} \\
 &  &  \nonumber \\
 & \ba{ccl}
   \dmmn(x) & \rightarrow  & e^{i\th(x)}\dmmn(x)   \\
   \emn(x)  & \rightarrow  & \emn(x)
   \ea
\qquad \qquad
   \ba{lcl}
   \mmm(x) & \rightarrow & e^{2i\th(x)} \mmm(x) \\
   \mm3(x) & \rightarrow & e^{i\th(x)} \mm3(x) \\
   \mmp(x) & \rightarrow & \mmp(x)    \\
   \md3(x) & \rightarrow & \md3(x),
   \ea  &     \label{sosu2transf}
\eeasn
i.e., the non-linear $SU(2)$ transformation is implemented by a
$U(1)_{local}$ transformation. The second  transformation 
in (\ref{asu2transf}a) allows us to introduce
a covariant derivative $D_{\m}~\equiv \pa_{\m}~\pm~i\bm$ acting on
$a_{\m}^{\pm}$ . Covariant derivatives acting on $d$s or $m$s 
are redundant and should not be considered (see \cite{Roman}).

The space-time transformations are given by
\beasn
& \phantom{,I}   \xi:\{ C_{3z}^+,\s_y \}:  &
      \left\{ \ba{ccc}
      a_{\m}^a   & \rightarrow & a_{\xi\m}^a       \\
      d_{pq}^a & \rightarrow & d_{\xi p \xi q}^a \\
      m^{ab}     & \rightarrow & m^{ab}
      \ea \right.                  \\
&               &   \nonumber         \\
& \phantom{C_{3z}^+,\s_y,} \xi: \{ I \}: &
      \left\{ \ba{ccc}
      a_{\m}^a   & \rightarrow & -a_{\xi\m}^{\bar a}  \\
      d_{pq}^a & \rightarrow & -d_{pq}^{\bar a} \\
      m^{ab}     & \rightarrow & m^{\bar a \bar b}
      \ea \right.                  \\
&               &   \nonumber         \\
& \phantom{\xi: \{ C_{3z}^+,\s_y,\} } T: &
      \left\{ \ba{ccc}
      a_{\m}^a   & \rightarrow & -a_{t\m}^{\bar a}  \\
      d_{pq}^a & \rightarrow & -d_{pq}^{\bar a} \\
      m^{ab}     & \rightarrow & m^{\bar a \bar b},
      \ea \right.
\label{aspacetransf}
\eeasn
where the symbols $\xi \m$, $\xi p$ and $t \m$ represents the transformation 
of the
subindex under the space and time transformations respectively together
with the corresponding coefficient in each case; the $\bar a$
superindex is the complex conjugate of $a$.

Next we present the way of introducing the coupling to
the electromagnetic field.
Since spin waves have no electric charge they couple to the
electromagnetic field through the field strength tensor, i.e., direct
couplings to the electric and magnetic fields. This kind of couplings
does not break the $SU(2)$ symmetry and in order to maintain the
space-time symmetry we impose the field ${\bf E}$ transforms like a
vector and the field ${\bf B}$ transforms like a pseudovector under
$\bar3m$ point group, whereas under time reversal these fields transform
as:
\be
   \ba{rl}
   T: &
      \left\{ \ba{ccc}
      E^a  & \rightarrow  &  E^a  \\
      B^a  & \rightarrow  &  -B^a.
      \ea \right.
   \ea
\label{fieldtimetransf}
\ee

Since the spin waves are fluctuations of magnetic moments there exists
another kind of coupling given by the Pauli term. The Pauli term breaks
explicitly the $SU(2)$ symmetry and a source with appropriate
transformation properties must be constructed to implement its effect in
the effective theory. In the Heisenberg lagrangian with the Pauli
interaction, written in the second quantisation language,
\be
L = \sum_i \psi^{\da}(x_i)
        i \left(\pa_0 -i\m\Sv{\bf B}(x_i)\right) \psi(x_i) + \cdots,
\ee
a source $A_0(x) \sim \m \Sv {\bf B}(x)$ can be associated to the Pauli
term, which
transforms like a connexion under time dependent $SU(2)$
transformations,
\be
A_0(x) \rightarrow g(t)A_0(x)g^{\dagger}(t)
                    + ig(t)\partial_0g^{\dagger}(t),
\ee
such that now the theory will be invariant under time dependent $SU(2)$
transformations. Therefore the effect of the Pauli term is implemented
in the
effective theory by changing the time derivative by a covariant time
derivative,
\be
\partial_0 \rightarrow D_0 \equiv \partial_0 -iA_0(x),
\ee
and eventually setting $A_0(x) = \m \Sv {\bf B}(x)$.
Once this change is performed one has to keep in mind that
the $a_0^{\pm}$ and $a_0^3$ contain the
magnetic field encoded in the covariant time derivative.

At this point the two sources of electromagnetic coupling to spin waves
have been considered.

%%%%%%%%%%%%%%%%% EFFECTVE LAGRANGIAN %%%%%%%%%%%%%%%%%%%%%%%%%%%%%%%%

\section{Effective Lagrangian}
\indent

Now we are in a position to construct the spin wave interaction
with the electromagnetic field for the antiferromagnet.
The way we choose to do this is a perturbative one: the derivative
expansion. Carrying out this expansion to a given order is meaningful
for
low energy and momentum with respect to the typical scales of the
antiferromagnet, given respectively
by the superexchange constant $J \sim 10meV$ and the inverse of the
lattice parameter $1/a \sim 0.1{\AA}^{-1}$ (the velocity of propagation
of the spin waves relates both parameters, $v = Ja$. It has the
following value in $Cr_2 O_3$: $v \sim 10^{-4}c$ \cite{LB}). The
characteristic energy and momentum of the system are given by the
external inputs of the electromagnetic fields, which are the same,
namely, $\omega$, and therefore the space derivative is highly
suppressed
with respect to the time derivative, $v\pa_i \sim 10^{-4}\pa_0$. The
suppression of the spin-orbit tensor has already been given, $D \sim
10^{-2}J$ and $M \sim 10^{-4}J$. Terms proportional to $D^2$ and $M$
force the local magnetisation to be in the third direction an give rise
to an energy gap $\sim 10^{-2}J$ for the spin waves. The amplitude of the
electromagnetic field must be constrained for the
expansion to make sense. First, we consider the Pauli term. Since it is
associated to the time derivative it is suppressed by $J$.
We will assume that the remaining couplings of the electromagnetic field  
come from vector and scalar potential minimal couplings in a micros\-copic 
model. The former
is associated to a link and hence suppressed by $1/ea$ whereas the
latter is associated to a time derivative and hence suppressed by $J/e$.
Therefore the electric field will be
suppressed by $J/ea$, whereas the magnetic field
will be suppressed by $1/ea^2$. The microscopic model may also have
non-minimal couplings to the electromagnetic field arising from the
integration of higher scales of energy and momentum. These terms would be
suppressed by the above mentioned higher scales and will be neglected.
In any case, as far as they respect the $SU(2)$ and crystal point group
symmetries their only effect is to slightly modify the value
of the constants in the effective lagrangian, which are anyway unknown.

Any effect due to spin waves is expected to be enhanced when we approach
their energy gap. This is why we shall choose the energy of the
electromagnetic wave of that order of magnitude. When in addition the
amplitude of the electromagnetic wave is tunned so that $E \sim \pa_0$
the following relative suppressions hold:
\be
   \ba{ccc}
   \pa_0, eaE, d  & \sim  & 10^{-2}J   \\
   m              & \sim  & 10^{-4}J   \\
   \m B           & \sim  & 10^{-5}J   \\
   v\pa_i, eavB   & \sim  & 10^{-6}J.
   \ea
\label{suppressions}
\ee

Once the above relation are given we are prepared to construct the
relevant effective lagrangian, invariant under $SU(2)$ and space-time
transformations given by (\ref{asu2transf}), (\ref{aspacetransf}) and
(\ref{fieldtimetransf}), for the effect we want to study: Non reciprocal
effects in SHG and GB.

The effective lagrangian at the lowest order in which
electromagnetic field appears reads
\bea
S[\pi,E,B] & = & \int{dx f_{\pi}^2 \left\{
                             \apo\amo \right. } \nonumber \\
 & & \phantom{ \int{dx f_{\pi}^2 \left\{ \right.}} \mbox{} +
      Z_1 (\dpzz \dmzbzb + \dmzz \dpzbzb)  \nonumber \\
 & & \phantom{ \int{dx f_{\pi}^2 \left\{ \right.}} \mbox{} +
      Z_2 [(\dpz3 \dmzz + \dmz3 \dpzz) +
            (\dmzb3 \dpzbzb + \dpzb3 \dmzbzb)]  \nonumber \\
 & & \phantom{ \int{dx f_{\pi}^2 \left\{ \right.}} \mbox{} +
      Z_3 (\dpz3 \dmzb3 + \dmz3 \dpzb3)  \nonumber  \\
 & & \phantom{ \int{dx f_{\pi}^2 \left\{ \right.}} \mbox{} +
      Z_4 \ezz \ezbzb   \nonumber  \\
 & & \phantom{ \int{dx f_{\pi}^2 \left\{ \right.}} \mbox{} +
      Z_5 (\ez3 \ezz + \ezb3 \ezbzb)  \nonumber  \\
 & & \phantom{ \int{dx f_{\pi}^2 \left\{ \right.}} \mbox{} +
      Z_6 \ez3 \ezb3                                  \label{SW_action}   \\
 & & \phantom{ \int{dx f_{\pi}^2 \left\{ \right.}} \mbox{} +
      Z_7 \mmp  \nonumber  \\
 & & \phantom{ \int{dx f_{\pi}^2 \left\{ \right.}} \mbox{} +
      Z_8 \md3  \nonumber  \\
 & & \phantom{ \int{dx f_{\pi}^2 \left\{ \right.}} \mbox{} +
      Z_9 i[(\dpzbzb \amo - \dmzbzb \apo)E^z -
               (\dmzz \apo - \dpzz \amo)E^{\zb}] \nn \\
 & & \phantom{ \int{dx f_{\pi}^2 \left\{ \right.}} \mbox{} +
      Z_{10} i[(\dpz3 \amo - \dmz3 \apo)E^z -
                 (\dmzb3 \apo - \dpzb3 \amo)E^{\zb}]  \nn \\
 & & \phantom{ \int{dx f_{\pi}^2 \left\{ \right.}} \mbox{} +
      Z_{11} E^zE^{\zb}  \nonumber \\
 & & \phantom{ \int{dx f_{\pi}^2 \left\{ \right.}} \mbox{} +
      \left. Z_{12} E^3E^3 \right\}, \nonumber
\eea

When we take into account the Pauli coupling in $a_0^{\pm}$, contributions
to SHG arise as $10^{-11}$ and $10^{-12}$ effects, and contributions to
$10^{-9}$ appear in the case of GB. As it will be shown these
contributions give rise to the desired non-reciprocal effects. Its is important
to notice that if a primitive translation mapping points with opposite
magnetisation existed the terms with a single time derivative above would not 
appear in the effective lagrangian.

Our action has been constructed up to third order
($10^{-6}$) and the contributions to non-reciprocal effects arise only
from the Pauli coupling which is much more suppressed in
(\ref{suppressions}). Therefore we might expect other
contributions at higher orders. This is indeed the case, but in order to
keep manegable the number of terms in the main text, these remaining contributions
are relegated to the appendix.

%%%%%%%%%%%%%%% E.M. FIELD EFFECTIVE INTERACTION %%%%%%%%%%%%%%%%%%%%%

\section{Electromagnetic Field Effective Interaction}
\indent

Our purpose is to describe non-reciprocal effects in SHG and GB mediated
by spin waves. Spin waves are responsible for an effective interaction
of the electromagnetic field giving rise to susceptibility tensors where
the properties of the material (spin waves) are encoded.

Hence we realise that spin waves are not to be observed in this
experiments and
therefore they must be eliminated from our theory. The way to do this is
by integrating them out in the functional (path) integral \cite{pi} 
so that the new action
depends only on the electromagnetic field. In order to perform the
integration we have to write the action explicitly in terms of the spin
waves. This is achieved by expanding (\ref{spinwaves}), with the
following result:
\bea
S[\pi,E,B] & = & \int{ dx \bigg[
  \pa_0\pi^+ \pa_0\pi^- - \Delta^2\pi^+\pi^-  } \nonumber \\
& & \phantom{\int{dx \left\{ \right.}} \mbox{} +
    i\m (\pi^+\pa_0\pi^- - \pi^-\pa_0\pi^+) B^3 \nonumber \\
& & \phantom{\int{dx \left\{ \right.}} \mbox{} -
    {1 \over 2} f_{\pi}[\pa_0\pi^+(\m B^{\zb} + \l E^{\zb}) +
                             \pa_0\pi^-(\m B^z + \l E^z)] \nonumber \\
& & \phantom{\int{dx \left\{ \right.}} \mbox{} -
    {1 \over 2}i\m f_{\pi}  [\pi^+(\m B^{\zb} + \l E^{\zb}) -
                          \pi^-(\m B^z + \l E^z)] B^3     \label{expansion}  \\
& & \phantom{\int{dx \left\{ \right.}} \mbox{} +
    {1 \over 4} \m \l f_{\pi}^2  (E^zB^{\zb} + E^{\zb}B^z)    \nn \\
& & \phantom{\int{dx \left\{ \right.}} \mbox{} +
    f_{\pi}^2\left(b_1 E^zE^{\zb} + b_2 E^3E^3\right) \bigg], \nonumber 
\eea
where only the terms contributing to bilinear and trilinear
electromagnetic fields in the effective action to be calculated are
kept. These terms are the only ones needed to describe
the desired effects.
The new constants which appear in (\ref{expansion}) are combinations of
those in the previous section. Although we do not know their precise
values, their order of magnitude is fixed according to the counting
rules given in (\ref{suppressions}). We list them below:
\bea
f_{\p}^2 \sim {1\over Ja^3} & & \Delta \sim D  \nn \\
b_{i}\sim (ea)^2            & & \l \sim ea{D\over J}.
\eea
Recall that $D\sim 10^{-2}J$ stands for the size of the spin-orbit
term.

%\newpage

Notice that the contributions to non-reciprocal effects
to the leading order ($10^{-11}$ and $10^{-12}$ for SHG and $10^{-9}$
for GB) come from terms with at most two spin waves, which permits us
to perform a gaussian integration in the functional generator. In
addition to this, it is worth mentioning that at the order given above
the
effects are produced at tree level, i.e., without loop contributions.

Once the gaussian integration is carried out a perturbative expansion
of the spin waves propagator in the presence of electromagnetic fields
has to be made, considering the free spin waves propagator,
\be
P(x-y) = \int{{dq \over (2\pi)^4} P(\o) e^{-iq(x-y)}}
\qquad ,\qquad
P(\o) = {1 \over \omega^2 - \Delta^2},
\label{propagator}
\ee
as the unperturbed part, leading to the electromagnetic effective
interaction lagrangian
\bea
S_{{\it eff}}[E,B] & = &
\int{ dx f_{\pi}^2 \left[ b_1 E^z E^{\zb} + b_2 E^3 E^3
        + {1 \over 4} \m\l (E^zB^{\zb} + E^{\zb}B^z) \right] } \nn \\
& + & \int{ dx dy f_{\pi}^2 \left[ {1 \over 4}\m\l
 \left(E^z \pa_0^2P(x-y) B^{\zb} + B^z \pa_0^2P(x-y)E^{\zb} \right) \right]}
                                                         \nn  \\
& + & \int{ dx dy f_{\pi}^2 \Bigg[
 - {1 \over 4} i\m \l^2
      E^z \left( B^3\pa_0P(x-y) + \pa_0P(x-y)B^3 \right) E^{\zb}} \label{emint} \\
& & \phantom{\int{ dx dy f_{\pi}^2 \Bigg\{ }}
    - {1 \over 4} i\m^2 \l
     \left[ E^z \left( B^3\pa_0P(x-y) + \pa_0P(x-y)B^3 \right) B^{\zb}
                                                  \right.     \nn \\
& & \phantom{\int{ dx dy f_{\pi}^2 \Bigg\{
                  - {1 \over 4} i\m^2 \l \left[ \right.} } + \left.
   B^z \left( B^3\pa_0P(x-y) + \pa_0P(x-y)B^3 \right) E^{\zb} \right]\Bigg].
                                                               \nn  
\eea 
The arguments of the electromagnetic fields have not been explicitly 
displayed. They must be understood as the nearest in the closest propagator.

Given the transformations under time reversal (\ref{fieldtimetransf}) it is
clear that the SHG, terms with three fields, presents non-reciprocal
effects due to the interference of different terms. The same is true for
the GB since bilinear terms proportional to the magnetic and electric
field appear.

 From the action above, together with the additional terms given in 
(\ref{emint_2}), the electromagnetic response of the $Cr_2 O_3$ due to spin waves, 
leading to non-reciprocal effects in SHG and GB, can be easily obtained using
\mbox{(\ref{pm_GB})-(\ref{pm_der})}.

The linear susceptibilities read
\bea
\x_E^{z\zb}(\o) & = & 2b_1 f_{\pi}^2  \nn  \\
\x_B^{z\zb}(\o) & = & {1 \over 2}\m\l f_{\pi}^2 [1 - \o^2P(\o)] \nn \\
\x_E^{33}(\o)   & = & 2b_2 f_{\pi}^2 \label{GB} \\
                &   &             \nn \\
\g_E^{z\zb}(\o) & = & {1 \over 2}\m\l f_{\pi}^2 [1 - \o^2P(\o)]. \nn
\eea
Recall that $ \x_B^{z\zb}(\o)$ and $\g_E^{z\zb}(\o)$ give rise to
the GB. Notice that here this effect is proportional to the gap of
the spin
wave spectrum, which is in turn due to the spin-orbit interaction.

The susceptibilities contributing to SHG read
\bea
\x_{EE}^{zzz}(\o,\o) & = &
     \l e_1f_{\pi}^2 \o^3[P(\o) - 4P(2\o)]
     + 2\l f_1f_{\pi}^2 \o[P(\o) - P(2\o)]        \nn  \\
& & \mbox{} + {1 \over 2} \l^2 d_1f_{\pi}^2 \o^3 P(\o)[P(\o) + 2P(2\o)]
                                             \nn \\
\x_{EE}^{z\zb 3}(\o,\o) & = &
   {1 \over 2} \l e_2f_{\pi}^2 \o^3[P(\o) + 4P(2\o)]
   - \l e_3f_{\pi}^2 \o^3[P(\o) - 2P(2\o)]      \nn \\
& & \mbox{} - {1 \over 2} \l f_2f_{\pi}^2 \o[P(\o) + 2P(2\o)]
      - {3 \over 2} \l^2 d_2f_{\pi}^2 \o^3[P(\o)P(2\o)] \nn  \\
& & \mbox{} - 3 \l^2 c f_{\pi}^2 \o^5 P(\o)P(2\o) - 3hf_{\pi}^2\o \nn \\
\x_{EB}^{zzz}(\o,\o) & = &
  \l g_1 f_{\pi}^2 \o [P(\o) - 2P(2\o)]   \label{SHG} \\
\x_{EB}^{z\zb 3}(\o,\o) & = &
  {1 \over 2}\m \l^2 f_{\pi}^2 \o [P(\o) + 2P(2\o) - 6\o^2P(\o)P(2\o)]
                                \nn \\
& & \mbox{} - \l g_2 f_{\pi}^2\o [P(\o) + 2P(2\o)]
                                  - 6 j_1 f_{\pi}^2\o  \nn \\
\x_{EB}^{z3\zb}(\o,\o) & = &
  -2\l g_3 f_{\pi}^2 \o P(2\o) + 2 j_2 f_{\pi}^2 \o \nn \\
\x_{BB}^{z\zb 3}(\o,\o) & = &
  {1 \over 4}\m^2 \l f_{\pi}^2 \o [P(\o) + 2P(2\o) - 6\o^2P(\o)P(2\o)]
                                                    \nn \\
\x_{EE}^{3\zb z}(\o,\o) & = &
 - \l e_2f_{\pi}^2 \o^3P(\o) + {1 \over 2} \l e_3f_{\pi}^2\o^3P(\o)\nn\\
\x_{EB}^{3\zb z}(\o,\o) & = &
- {1 \over 2} \l g_3 f_{\pi}^2 \o P(\o) - 2j_2 f_{\pi}^2 \o  \nn \\
                        &   &                       \nn \\
\g_{EE}^{zzz}(\o,\o) & = & \l g_1 f_{\pi}^2 \o P(\o)  \nn \\
\g_{EE}^{z\zb 3}(\o,\o) & = &
-{1 \over 2} \l g_3 f_{\pi}^2 \o P(\o) + j_2 f_{\pi}^2 \o   \nn \\
\g_{EB}^{z\zb 3}(\o,\o) & = &
 {1 \over 2}\m^2 \l f_{\pi}^2 \o [P(\o) + 2P(2\o) - 6\o^2P(\o)P(2\o)].
\nn
\eea

Notice the non-trivial dependence in $\o$ of the above susceptibilities.
(This dependence is slightly more involved if one calculates the general
susceptibilities $\x (\o ,\o^{\prime})$ and $\g(\o ,\o^{\prime})$ since
the limit $\o = \o^{\prime}$ produces a few cancellations.) For this to be so it is
crucial that no primitive translation mapping points with opposite magnetisation
exists. Otherwise only the local terms proportional to $j_{i}$ would survive.
Notice also that the terms proportional to $h$ arise due to the explicit 
spin-orbit breaking. Even thought this term gives rise to a local term
in the electromagnetic fields at this order, it also contains explicit
interactions with spin waves at higher orders.

Let us finally mention that the terms proportional to $k_{i }$ in
(\ref{emint_2}) give rise to contributions to the quadrupolar momentum 
\cite{Agranovich,Pershan} which are of the
same order as the ones considered above. In fact, the quadrupolar terms
give the unique contribution to SHG in a crystal which contains a center
of symmetry as it is the case for $Cr_2O_3$ above the Neel temperature
\cite{Pershan}. The associated susceptibilities
can be easily calculated. They are local and will not be displayed
explicitly.

%%%%%%%%%%%%%%%%%%% DISCUSSION %%%%%%%%%%%%%%%%%%%%%%%%%%%%%%%%%%%%%%

\section{Discussion}
\indent

We have used an effective field theory for spin waves in an antiferromagnetic 
material to describe its response to electromagnetic fields in the microwave 
region. The starting point is a local effective lagrangian which fully exploits 
the fact that spin waves are Goldstone modes of a $SU(2) \rightarrow U(1)$ 
symmetry breaking pattern together with the crystal space group symmetry and time 
reversal. By integrating out the spin waves we obtain a non-local effective action 
which encodes the response of the material to the electromagnetic field. From this 
effective action the various linear and non-linear electric and magnetic 
susceptibilities can be immediately obtained. We have given explicitly those 
relevant to the GB and SHG experiments. These susceptibilities depend on a 
relatively large number of unknown constants ($\sim 23$) and a microscopic 
calculation is required to assign definite numbers to them. However, their order of
magnitude can be readily established in terms of the typical lattice spacing $a$
and the energy of the first gapped excitation $J$. Notice also that these
susceptibilities present a rather non-trivial dependence on $\omega$, the 
frequency of the incoming radiation. This dependence cannot be obtained from the 
magnetic group symmetries alone and it is a direct consequence of the existence of
spin waves in an antiferromagnetic crystal where: (i) no primitive translation 
mapping points with opposite magnetisations exist, and (ii) spin-orbit effects 
are sizable.

As mentioned in the introduction, from group theoretical arguments it has been 
known for long that certain
antiferromagnets may support non-reciprocal effects \cite{Birss,theoretical}.
 However, group theoretical
considerations alone are unable to indicate any mechanism by which these 
effects may be realised, not even to provide an order of magnitude estimate. In 
ref.~\cite{experimental} non-reciprocal effects were observed in $Cr_2 O_3$ in the 
optical region, and in ref.~\cite{Valenti} a theoretical explanation was presented. 
A microscopic mechanism leading to such effects 
was identified in atomic transitions in which spin-orbit interactions
and the trigonal field play a crucial r\^ole. Here, we have presented a totally
different mechanism which also leads to non-reciprocal effects but in the microwave
region. Namely, the
interaction of spin waves with microwave radiation. Since the mechanisms are
completely different, the susceptibilities given in \cite{Valenti} hardly have any
resemblance with ours, except for the general group theoretical constraints that both
of them must fulfil. For instance, the susceptibilities in \cite{Valenti} tipically
depend on the size of the of the $Cr$ ions, the energy differences between atomic
levels, and matrix elements of perturbations (like spin-orbit terms) between atomic
states. Instead, ours tipically depend on the lattice spacing, the Heisenberg coupling
and the spin wave energy gap. However, some common features do exist. The spin wave
energy gap is due to spin-orbit terms, which are crucial to obtain
non-reciprocal effects both for ref.~\cite{Valenti} and for us. In fact, within our
approach it is clear that spin-orbit terms are the responsible for the spins to point
to the third direction, and hence for the magnetic group to be what it is.
Also the magnetoelectric susceptibilities in ref.~\cite{Valenti} are proportional to the
magnetic moment of the ion and so are ours. Nevertheless, for the optical region it
seems that only the susceptiblities $\g_E$, $\chi_{EE}$ and $\g_{EE}$ are relevant 
and, then, it is crucial that $\chi_{EE} \sim\g_{EE} \not= 0$ for the observation of
non-reciprocal effects in second harmonic generation.
In our case all the susceptibilities (i.e. including
$\chi_{EB}$, $\chi_{BB}$, $\g_{EB}$ and $\g_{BB}$)
have the same order of magnitude which provides
further observational possibilities. For the sake of comparison, we give below our
order of magnitude estimates for the susceptibilities given in \cite{Valenti}, assuming
the orders of magnitude that we have been using for the parameters so far. From our
$\g_{E}$, $\chi_{EE}$ and $\g_{EE}$ we find, adopting the notation of   
\cite{Valenti}, $\a^{xx}\sim 10^{-2}$ and $\chi\sim \g\sim 10^{-10} CN^{-1}$, which
appear to be a few orders of magnitude larger.

We have not been able to locate experimental results in the literature to test our
formulas against. We expect them to become available at some point. It would be 
particularly interesting to be able to browse the microwave region with several 
frequencies so that the $\omega$ dependence in (\ref{GB}) and (\ref{SHG}) could be 
checked and the free parameters fitted. If the incoming radiation is directed 
along the third axis then only $\g_{EE}^{zzz}(\o,\o)$, $\x_{EB}^{3\zb z}(\o,\o)$,
$\x_{EE}^{3\zb z}(\o,\o)$, $\x_{EB}^{zzz}(\o,\o)$ and $\x_{EE}^{zzz}(\o,\o)$ are 
relevant.

Although for definiteness we have focused on the $Cr_2O_3$ crystal, which has spin 
$3/2$, our results hold for any antiferromagnetic crystal (with spontaneous 
staggered magnetisation) with crystal point group $\bar 3 m$ and arbitrary spin, 
as long as no primitive translations mapping points with opposite magnetisation 
exist. This includes for instance $V_2 O_3$ (spin~$1$). It is also worth
emphasising that the method we have used is general enough to become applicable to 
any antiferromagnet of any spin and crystal point group, as long as there is 
spontaneous staggered magnetisation. The allowed terms in the effective lagrangian,
however, depend on the particular crystal point group and on the particular 
distribution of the magnetic ions in the crystal.

%\newpage

%%%%%%%%%%%%%%%%%%%%%%%% ACKNOWLEDGEMENTS %%%%%%%%%%%%%%%%%%%%%%%%%%%%%%

\section*{Acknowledgements}
\indent

We thank P.~Hasenfratz, F.~Niedermayer and J.~L.~Ma\~nes for useful
conversations. We
also thank R. Valent\'{\i} for explanations on \cite{Valenti}.
J.~M.~R.~thanks J.~Llum\`a, M.~Garc\'{\i}a del Muro, J.~M.~Ruiz,
O.~Iglesias and A.~Labarta,
in the magnetism group, for helpful comments. He is supported by a
Basque Goverment F.P.I.~grant. Financial
support from CICYT, contract AEN95-0590 and from CIRIT, contract
GRQ93-1047 is also acknowledged.

%%%%%%%%%%%%%%%%%%%%%%%%%%%%%%%%%%%%%%%%%%%%%%%%%%%%%%%%%%%%%%%%%%%%%%%

\appendix

\Appendix{}
\indent

In this appendix we will present the higher order terms in the effective action (4.2) 
which contribute to he SHG and GB to the same order as those in (\ref{expansion}).

Contributions to fourth order $(10^{-8})$:
\beasn
& & i(\apo D_0\amo - \amo D_0\apo)E^3              \\
& &                                     \nonumber  \\
& & \ezz(\dpzz\amo + \dmzz\apo)E^z +
         \ezbzb(\dmzbzb\apo + \dpzbzb\amo)E^{\zb}  \nonumber \\
& & \ezz(\dpzb3\amo + \dmzb3\apo)E^z +
         \ezbzb(\dmz3\apo + \dpz3\amo)E^{\zb}         \nn \\
& & \ezb3(\dpzb3\amo + \dmzb3\apo)E^z +
         \ez3(\dmz3\apo + \dpz3\amo)E^{\zb} \nn \\
%& &                                      \nonumber  \\
& & [\ezz(\dpzbzb\amo + \dmzbzb\apo) +
          \ezbzb(\dmzz\apo + \dpzz\amo)]E^3         \\
& & [\ezz(\dpz3\amo + \dmz3\apo) +
          \ezbzb(\dmzb3\apo + \dpzb3\amo)]E^3         \nn \\
& & [\ezb3(\dpz3\amo + \dmz3\apo) +
           \ez3(\dmzb3\apo + \dpzb3\amo)]E^3.  \nn
\label{additional_8}
\eeasn

%\newpage

To fifth order $(10^{-10})$:
\beasn
& & (\dpzz D_0\amo + \dmzz D_0\apo)E^zE^z +
          (\dmzbzb D_0\apo + \dpzbzb D_0\amo)E^{\zb}E^{\zb}  \nn \\
& & (\dpzb3 D_0\amo + \dmzb3 D_0\apo)E^zE^z +
          (\dmz3 D_0\apo + \dpz3 D_0\amo)E^{\zb}E^{\zb}       \nn\\
& & (\dpzz D_0\amo + \dmzz D_0\apo)E^{\zb}E^3 +
          (\dmzbzb D_0\apo + \dpzbzb D_0\amo)E^zE^3     \nn \\
& & (\dpzb3 D_0\amo + \dmzb3 D_0\apo)E^{\zb}E^3 +
          (\dmz3 D_0\apo + \dpz3 D_0\amo)E^zE^3         \\
& & (\dpzz \amo + \dmzz \apo)E^{\zb}\pa_0 E^3 +
          (\dmzbzb \apo + \dpzbzb \amo)E^z\pa_0 E^3      \nn \\
& & (\dpzb3 \amo + \dmzb3 \apo)E^{\zb}\pa_0 E^3 +
          (\dmz3 \apo + \dpz3 \amo)E^z\pa_0 E^3     \nn \\
& &          \nonumber  \\
%%%%% Two white lines to pass the page
%& &          \nonumber  \\
%& &          \nonumber  \\
& & i(\dpzz\dmzbzb - \dmzz\dpzbzb)(\ezz E^zE^z - \ezbzb E^{\zb}E^{\zb})
                                                           \nn \\
& & i(\dpz3\dmzb3 - \dmz3\dpzb3)(\ezz E^zE^z - \ezbzb E^{\zb}E^{\zb})
                                                            \nn \\
& & i(\dpzz\dmzbzb - \dmzz\dpzbzb)(\ezb3 E^zE^z - \ez3 E^{\zb}E^{\zb})
                                                            \nn \\
& & i(\dpz3\dmzb3 - \dmz3\dpzb3)(\ezb3 E^zE^z - \ez3 E^{\zb}E^{\zb})
                                                            \nn \\
& & i[(\mp3\dmzz -\mm3\dpzz)E^zE^z -
         (\mm3\dpzbzb -\mp3\dmzbzb)E^{\zb}E^{\zb}]   \nn \\
& & i[(\mp3\dmzb3 -\mm3\dpzb3)E^zE^z -
         (\mm3\dpz3 -\mp3\dmz3)E^{\zb}E^{\zb}]     \nn \\
%& &           \nonumber  \\
& & i(\dpzz\dmzbzb - \dmzz\dpzbzb)(\ezz E^{\zb}E^3 - \ezbzb E^zE^3)
                                                                \\
& & i(\dpz3\dmzb3 - \dmz3\dpzb3)(\ezz E^{\zb}E^3 - \ezbzb E^zE^3)
                                                            \nn \\
& & i(\dpzz\dmzbzb - \dmzz\dpzbzb)(\ezb3 E^{\zb}E^3 - \ez3 E^zE^3)
                                                            \nn \\
& & i(\dpz3\dmzb3 - \dmz3\dpzb3)(\ezb3 E^{\zb}E^3 - \ez3 E^zE^3)
                                                            \nn \\
& & i[(\mp3\dmzz -\mm3\dpzz)E^{\zb}E^3 -
         (\mm3\dpzbzb -\mp3\dmzbzb)E^zE^3]       \nn \\
& & i[(\mp3\dmzb3 -\mm3\dpzb3)E^{\zb}E^3 -
         (\mm3\dpz3 -\mp3\dmz3)E^zE^3]             \nn \\
& &             \nonumber  \\
& & i(\ezz E^zB^z - \ezbzb E^{\zb}B^{\zb})   \nn \\
& & i(\ezb3 E^zB^z - \ez3 E^{\zb}B^{\zb})    \nn \\
& & i(\ezz E^{\zb}B^3 - \ezbzb E^zB^3)       \nn \\
& & i(\ezb3 E^{\zb}B^3 - \ez3 E^zB^3)            \\
& & i(\ezz E^3B^{\zb} - \ezbzb E^3B^z)       \nn \\
& & i(\ezb3 E^3B^{\zb} - \ez3 E^3B^z).       \nn
\label{additional_10}
\eeasn

To sixth order $(10^{-12})$:
\beasn
& & i(\dpzz\dmzbzb - \dmzz\dpzbzb)(E^z\pa_0E^{\zb} -
                                  E^{\zb}\pa_0E^z)E^3  \nn \\
& & i[(\dpzz\dmz3 - \dmzz\dpz3)E^z\pa_0E^{\zb} -
      (\dmzbzb\dpzb3 - \dpzbzb\dmzb3)E^{\zb}\pa_0E^z]E^3   \\
& & i(\dpz3\dmzb3 - \dmz3\dpzb3)(E^z\pa_0E^{\zb} -
                                  E^{\zb}\pa_0E^z)E^3  \nn \\
& &       \nonumber  \\
& & (E^z\pa_0 E^{\zb} - E^{\zb}\pa_0 E^z)B^3     \nn \\
& & E^z\pa_0 E^3B^{\zb} - E^3\pa_0 E^{\zb}B^z        \\
& &       \nonumber   \\
& & (E^z\pa_{\zb}E^z + E^{\zb}\pa_z E^{\zb})E^3   \nn \\
& & E^zE^{\zb}\pa_3 E^3                           \nn \\
& & E^zE^{\zb}(\pa_z E^z + \pa_{\zb} E^{\zb})         \\
& & E^3E^3(\pa_z E^z + \pa_{\zb} E^{\zb}).        \nn
\label{additional_12}
\eeasn

Where to reduce the number of terms in (\ref{additional_12}b) the homogeneous Maxwell
equations, which are satisfied automatically, have been used.
It is important to notice that if a primitive translation mapping points with opposite
magnetisation existed, the terms in (\ref{additional_8}a), (\ref{additional_8}b), 
(\ref{additional_10}c) and (\ref{additional_12}a) would not appear.

In spite of the large number of terms, we will see that most of them 
contribute in the same way to the effective action for the electromagnetic 
fields.

Indeed, when we expand the terms, above in order to make explicit the interaction
between the spin waves and the electromagnetic field, the terms below must be added
to (\ref{expansion}) keeping bilinear and trilinear terms in the electromagnetic fields.
\bea
{\Delta S}[\pi,E,B] & = & \int{ dx \bigg[
    ic(\pa_0\pi^+\pa_0^2\pi^- - \pa_0\pi^-\pa_0^2\pi^+)E^3 }  \nn \\
& & \phantom{\int{dx \left\{ \right.}} \mbox{} +
    id_1(\pi^+\pa_0\pi^+E^z - \pi^-\pa_0\pi^-E^{\zb})  \nn \\
& & \phantom{\int{dx \left\{ \right.}} \mbox{} +
    id_2(\pi^+\pa_0\pi^-E^3 - \pi^-\pa_0\pi^+E^3)     \nn \\
& & \phantom{\int{dx \left\{ \right.}} \mbox{} +
    ie_1f_{\pi}(\pa_0\pi^+E^z\pa_0E^z -
                          \pa_0\pi^-E^{\zb}\pa_0E^{\zb}) \nonumber \\
& & \phantom{\int{dx \left\{ \right.}} \mbox{} +
    ie_2f_{\pi}(\pa_0\pi^+\pa_0E^{\zb}E^3 -
                          \pa_0\pi^-\pa_0E^zE^3) \nonumber \\
& & \phantom{\int{dx \left\{ \right.}} \mbox{} +
    ie_3f_{\pi}(\pa_0\pi^+E^{\zb}\pa_0E^3 -
                          \pa_0\pi^-E^z\pa_0E^3)      \nn \\
& & \phantom{\int{dx \left\{ \right.}} \mbox{} +
    if_1f_{\pi}(\pi^+E^zE^z - \pi^-E^{\zb}E^{\zb})    \nn \\
& & \phantom{\int{dx \left\{ \right.}} \mbox{} +
    if_2f_{\pi}(\pi^+E^{\zb}E^3 - \pi^-E^zE^3)        \nn \\
& & \phantom{\int{dx \left\{ \right.}} \mbox{} +
    ig_1f_{\pi}(\pi^+E^zB^z - \pi^-E^{\zb}B^{\zb})     \nn \\
& & \phantom{\int{dx \left\{ \right.}} \mbox{} +
    ig_2f_{\pi}(\pi^+E^{\zb}B^3 - \pi^-E^zB^3)         \label{expanssion_2} \\
& & \phantom{\int{dx \left\{ \right.}} \mbox{} +
    ig_3f_{\pi}(\pi^+E^3B^{\zb} - \pi^-E^3B^z)            \nn \\
& & \phantom{\int{dx \left\{ \right.}} \mbox{} +
    ih f_{\pi}^2 (E^z\pa_0 E^{\zb} - E^{\zb}\pa_0 E^z)E^3  \nn \\
& & \phantom{\int{dx \left\{ \right.}} \mbox{} +
    ij_1 f_{\pi}^2 (E^z\pa_0 E^{\zb} - E^{\zb}\pa_0 E^z)B^3  \nn \\
& & \phantom{\int{dx \left\{ \right.}} \mbox{} +
    ij_2 f_{\pi}^2 (E^z\pa_0 E^3B^{\zb} - E^3\pa_0 E^{\zb}B^z) \nn \\
& & \phantom{\int{dx \left\{ \right.}} \mbox{} +
    k_1 f_{\pi}^2 (E^z\pa_{\zb}E^z + E^{\zb}\pa_zE^{\zb})E^3  \nn \\
& & \phantom{\int{dx \left\{ \right.}} \mbox{} +
    k_2 f_{\pi}^2 E^zE^{\zb}\pa_3 E^3  \nonumber  \\
& & \phantom{\int{dx \left\{ \right.}} \mbox{} +
    k_3 f_{\pi}^2 E^zE^{\zb}(\pa_z E^z + \pa_{\zb} E^{\zb})   \nn \\
& & \phantom{\int{dx \left\{ \right.}} \mbox{} +
    k_4 f_{\pi}^2 E^3E^3(\pa_z E^z + \pa_{\zb} E^{\zb}) \bigg]. \nn
\eea

Notice that no terms beyond two spin waves appear in the previous action. This 
does not change the procedure of gaussian integration carried out in section~5.
The constants in (\ref{expanssion_2}) have the following order of magnitude:

\bea
c \sim {ea \over J^2} & & d_{i} \sim ea\left({D\over J}\right)^2 \nn \\
e_{i} \sim \left({ea\over J}\right)^2{D\over J} & &
                  f_{i}\sim  (ea)^2\left({D\over J}\right)^3  \nn\\
g_{i}\sim (ea)^2 Da  & &  h\sim (ea)^3 D^2    \\
j_{i}\sim (ea)^3 {a\over J}  & &  k_{i}\sim (ea)^3 {a\over J}.  \nn
\eea

The effective action for the electromagnetic fields in (\ref{emint}) must be 
augmented with the following terms
{\small
\bea
\lefteqn{
{\Delta S}_{{\it eff}}[E,B] =
 \int{ dx dy f_{\pi}^2 \Bigg\{ - {1 \over 4} i \l e_1
       \left[ E^z \pa_0^3P(x-y) E^zE^z +
              E^{\zb}E^{\zb} \pa_0^3P(x-y) E^{\zb} \right] }} \nn \\
& & \mbox{} \phantom{ \qquad \qquad \!\! + \int{ dx dy f_{\pi}^2 \Bigg\{ }}
     - {1 \over 2} i \l
  \left[ E^z \pa_0^2P(x-y) (e_2 E^{\zb}\pa_0E^3 + e_3 \pa_0E^{\zb}E^3)
                                               \right.  \nn \\
& & \mbox{} \phantom{ \qquad \qquad \!\! + \int{ dx dy f_{\pi}^2 \Bigg\{
                     - {1 \over 2} i \l \left[ \right. } } + \left.
   (e_2 E^z\pa_0E^3 + e_3 \pa_0E^zE^3) \pa_0^2P(x-y) E^{\zb} \right]
                                                                 \nn \\
& & \mbox{} \phantom{ \qquad \qquad \!\! + \int{ dx dy f_{\pi}^2 \Bigg\{ }}
+ {1 \over 2} i \l
  \left[ E^z \pa_0 P(x-y) (f_1 E^zE^z + f_2 E^{\zb}E^3) \right. \nn\\
& & \mbox{} \phantom{ \qquad \qquad \!\! + \int{ dx dy f_{\pi}^2 \Bigg\{
                     + {1 \over 2} i \l \left[ \right. } } + \left.
    (f_1 E^{\zb}E^{\zb} + f_2 E^zE^3) \pa_0 P(x-y) E^{\zb} \right]
                                                      \nn \\
& & \mbox{} \phantom{ \qquad \qquad \!\! + \int{ dx dy f_{\pi}^2 \Bigg\{ }}
+ {1 \over 2} i \l
  \left[ E^z \pa_0 P(x-y) (g_1 E^zB^z + g_2 E^{\zb}B^3
                                       + g_3 E^3B^{\zb}) \right. \nn\\
& & \mbox{} \phantom{ \qquad \qquad \!\! + \int{ dx dy f_{\pi}^2 \Bigg\{
                     + {1 \over 2} i \l \left[ \right. } } + \left.
   (g_1 E^{\zb}B^{\zb} + g_2 E^zB^3 + g_3 E^3B^z)
\pa_0 P(x-y) E^{\zb} \right]    \Bigg\}                 \label{emint_2} \\
%%%%%%%%%%%%%
%& & \nn \\
%& & \nn \\
%%%%%%%%%%%%%
& & \mbox{} + \int{ dx du dy f_{\pi}^2 \Bigg\{
  - {1 \over 4}i \l^2
  E^z \Big(\pa_0^2P(x-u)(\m B^3 + d_2E^3)\pa_0P(u-y) }   \nn \\
& & \mbox{} \phantom{+ \int{ dx du dy f_{\pi}^2 \Bigg\{
                    - {1 \over 4}i \l^2 E^z \Big( } }
      +\pa_0P(x-u)(\m B^3 + d_2E^3)\pa_0^2P(u-y) \Big) E^{\zb}  \nn \\
& & \mbox{} \phantom{+ \int{ dx du dy f_{\pi}^2 \Bigg\{ }}
- {1 \over 4} i \m^2 \l \bigg[
  E^z \Big(\pa_0^2P(x-u)B^3\pa_0P(u-y)
       + \pa_0P(x-u)B^3\pa_0^2P(u-y) \Big) B^{\zb}  \nn \\
& & \mbox{} \phantom{+ \int{ dx du dy f_{\pi}^2 \Bigg\{
                     - {1 \over 4} i \m^2 \l \bigg[ }}
  + B^z \Big(\pa_0^2P(x-u)B^3\pa_0P(u-y)
         + \pa_0P(x-u)B^3\pa_0^2P(u-y) \Big) E^{\zb} \bigg]  \nn \\
& & \mbox{} \phantom{+ \int{ dx du dy f_{\pi}^2 \Bigg\{ }}
  + {1 \over 4}i \l^2 c
  E^z \Big( \pa_0^2P(x-u)E^3\pa_0^3P(u-y)
      + \pa_0^3P(x-u)E^3\pa_0^2P(u-y) \Big) E^{\zb}  \nn \\
& & \mbox{} \phantom{+ \int{ dx du dy f_{\pi}^2 \Bigg\{ }}
         + {1 \over 8} i d_1 \l^2 \bigg[
  E^z \Big(\pa_0^2P(x-u) E^z \pa_0P(u-y)
       - \pa_0P(x-u) E^z \pa_0^2P(u-y) \Big) E^z  \nn \\
& & \mbox{} \phantom{+ \int{ dx du dy f_{\pi}^2 \Bigg\{
                     + {1 \over 8} i d_1 \l^2 \bigg[ } }
  - E^{\zb} \Big( \pa_0^2P(x-u) E^{\zb} \pa_0P(u-y)
             - \pa_0P(x-u) E^{\zb} \pa_0^2P(u-y) \Big) E^{\zb} \bigg]
                                          \Bigg\}  \nn \\
%%%%%%
& & \mbox{} + \int{ dx f_{\pi}^2 \bigg[
       i h (E^z\pa_0E^{\zb} - E^{\zb}\pa_0E^z)E^3 }  \nn \\
& & \mbox{} \phantom{\int{ dx f_{\pi}^2 \left[ \right. } }
     + i j_1 (E^z\pa_0E^{\zb} - E^{\zb}\pa_0E^z)B^3
     + i j_2 (E^zB^{\zb} - E^{\zb}B^z) \pa_0E^3  \nn \\
& & \mbox{} \phantom{\int{ dx f_{\pi}^2 \left[ \right. } }
     + k_1 (E^z\pa_{\zb}E^z + E^{\zb}\pa_zE^{\zb}) E^3
     + k_2 E^zE^{\zb}\pa_3E^3   \nn \\
& & \mbox{} \phantom{\int{ dx f_{\pi}^2 \left[ \right. } }
     + k_3 E^zE^{\zb}(\pa_zE^z + \pa_{\zb}E^{\zb})
     + k_4 E^3E^3(\pa_zE^z + \pa_{\zb}E^{\zb}) \bigg]. \nn
\eea}

These contributions to the electromagnetic effective action have been included in our
final results for the generalised susceptibilities in formulas (\ref{GB}) and (\ref{SHG}).

%%%%%%%%%%%%%%%%%%%%%% BIBLIOGRAPHY %%%%%%%%%%%%%%%%%%%%%%%%%%%%%%%%%%%

\end{document}